\documentclass[aps,prb,twocolumn,showpacs,floatfix,superscriptaddress]{revtex4-1}
\usepackage[dvips]{graphicx}
\usepackage{dcolumn}
\usepackage{amsmath}
\usepackage{amssymb}
\usepackage{multirow}
\usepackage{color}
\usepackage{moreverb}
\usepackage{bm}
\usepackage{float}
\usepackage[english]{babel}

\begin{document}
\title{Designing Fe Nanostructures at Graphene/h-BN Interfaces}

\author{Soumyajyoti Haldar}
\affiliation{Department of Physics and Astronomy, Uppsala University, Box-516, SE 75120, Sweden}
\author{Pooja Srivastava}
\affiliation{Harish-Chandra Research Institute, Chhatnag Road, Jhunsi, Allahabad 211019, India}
\author{Olle Eriksson}
\affiliation{Department of Physics and Astronomy, Uppsala University, Box-516, SE 75120, Sweden}
\author{Prasenjit Sen}
\email{prasen@hri.res.in}
\affiliation{Harish-Chandra Research Institute, Chhatnag Road, Jhunsi, Allahabad 211019, India}
\author{Biplab Sanyal}
\email{Biplab.Sanyal@physics.uu.se}
\affiliation{Department of Physics and Astronomy, Uppsala University, Box-516, SE 75120, Sweden}

\begin{abstract}
Tailor-made magnetic nanostructures offer a variety of functionalities useful for technological applications. In this work, we explore the possibilities of realizing Fe nanostructures at the interfaces of 2D graphene and h-BN by ab initio density functional calculations. With the aid of ab initio Born-Oppenheimer molecular dynamics simulations and diffusion barriers calculated by nudged elastic band method, we find that (i) diffusion barriers of Fe on BN are much smaller than those on graphene, (ii) the Fe adatoms form clusters within a short time interval ($\sim$2.1 ps) and (iii) Fe clusters diffuse easily across the C-N interface but become immobile at the C-B interface. The calculated magnetic exchange coupling between Fe clusters at C-B interfaces varies non-monotonically as a function of the width of BN separating the graphene parts. One may envisage design of magnetic nanostructures at the C-B interface of 2D graphene/h-BN hybrids to realize interesting applications related to spintronics.
\end{abstract}

\maketitle

\section{Introduction}
The absence of a band gap makes use of graphene in field effect transistors
difficult~\cite{graphrev1} as one needs a high on-off ratio for the operation of a transistor. 
This has led to various attempts to open
a band gap in graphene by chemical functionalization or by creating nanostructures
in form of graphene nanoribbons. 
A parallel route has been adopted in building hybrid
materials involving graphene and hexagonal boron nitride (h-BN).
Hetero-structures of graphene and h-BN with alternating layers of
the two materials have been studied for the last few years~\cite{Giovannetti-07,Slawinska-10,Slawinska-10a,Haigh-12,Quhe-12} quite
extensively. Recently a lot of attention has been paid to hybrid hexagonal BNC (h-BNC) sheets in
which graphene and h-BN are combined in the same two dimensional
(2D) planes~\cite{Pruneda-10,Pruneda-12,bnc-ci,litho-bnc}. By altering the ratio of the two
materials it was possible to tune the gap of the material.
Theoretical calculations~\cite{Yuge-09,Zhu-11,Martins-11,Pooja-11} have
shown that the band gap depends on both the composition and the
exact arrangement of the C, B and N atoms. Pure graphene or h-BN sheets
have sp$^{2}$ hybrid electronic states and are nonmagnetic. Introduction of unpaired $d$ electrons
through transition metal adatoms gives rise to possibilities of finite magnetism, especially formation
of interesting and useful magnetic self-assemblies~\cite{carbone-11}. The formation of self-assemblies is related to the diffusion of atoms on graphene. The diffusion barriers of atoms on graphene have been calculated from first principles \cite{tim-09,sahin-12}. Not only graphene, but also h-BN have been used as a 2D template for studying energetics of adsorption and diffusion \cite{olegBN}. From the experimental side, the interaction between graphene and metals can be studied with high resolution scanning tunneling microscopy \cite{novoselovNL}.
Therefore, for most electronic applications, the use of these 2D materials  will involve
contacts with metals and hence, a fundamental understanding of the interaction between metals (adatoms,
clusters and thin films) and 2D graphene, h-BN or hybrid graphene/h-BN interfaces is
very important. 

Given this background, some attention has been paid to the adsorption of 3d transition metal (TM) atoms on h-BNC$_2$ (1:1 composition of graphene and h-BN) sheets. In one of these 
earlier works~\cite{Pooja-11} it was shown that Fe, Co and Ni are weakly adsorbed on the h-BNC$_2$ sheet.
Moreover, it was found that for a 32-atom supercell the hexagonal site within the graphene region is most favorable for a Fe adatom, but for larger
supercells the hexagonal site at the C-B interface becomes the most favorable one. 
Calculated diffusion barriers along a limited number of paths were rather small for
these adatoms. Weak adsorption and small energy barriers indicated that the adatoms
would be mobile on the sheets. This posed the question
whether it was energetically favorable for the adatoms to remain isolated or to form clusters on the h-BNC$_2$ sheet. In a simple picture, in which
the reference states of the TM atoms were taken as their respective bulk phases, 
we found the formation energies of the isolated adatoms on the sheet to be 
positive.
This indicated that it was energetically favorable for the adatoms to cluster together.

While the above work provided some insights into the behavior of TM impurities on the h-BNC$_2$ sheet, a lot of details about the system still remain
unanswered, e.g., the possibility of formation of clusters of TM atoms and their mobility
on the hybrid interfaces, magnetic properties of clusters etc. 
In this work we have studied in detail adsorption and diffusion of a single and multiple Fe adatoms and their magnetic properties 
on a h-BNC$_2$ sheet. In particular, we have calculated diffusion barriers for a single Fe adatom along all possible diffusion paths on 
different parts of the sheet. In order to test whether the calculated  diffusion
barriers can provide realistic estimates for the diffusion behavior of the Fe adatom 
at finite temperatures, we have performed finite temperature ab initio 
molecular dynamics simulations.
At room temperature, a single Fe adatom finds its equilibrium position at a 
hexagonal site at the C-B interface, a site that was found to be 
energetically most 
favorable in terms of the calculated adsorption energies. This prompted us to calculate the exchange interaction
between the magnetic moments on two different Fe atoms adsorbed at similar sites along the C-B interface. We then addressed the questions whether and how 
several Fe atoms deposited together on the sheet form clusters, and how they diffuse subsequently. Our finite temperature molecular dynamics (MD) 
calculations suggest that individual Fe adatoms diffuse rather fast and form clusters. The clusters diffuse rather smoothly across h-BN and graphene
regions, and the C-N interface, but eventually get trapped at the C-B interface. 
Finally we discuss the possibility of the formation of self-assembled magnetic nano-structures in this system.

\section{Methods}
All our first-principles calculations were performed using a plane wave basis 
within density functional theory as implemented in the VASP code \cite{vasp}. An 
energy cutoff of 550 eV was used to expand the wave functions in our zero 
temperature calculations of force minimizations.
Generalized gradient approximation as proposed by Perdew, Burke and 
Ernzerhof~\cite{PBE,PBEerr}  was used for the exchange-correlation functional. 
The effect of the core 
electrons were removed by using the projected augmented wave (PAW) 
potentials~\cite{blo,blo1}. The structures were optimized using the conjugate 
gradient method
with forces calculated from the Hellman-Feynman theorem. Structures were 
considered to have been optimized when all the forces were smaller than 0.01 
eV/\AA.
Diffusion barriers for a single Fe adatom on a h-BNC$_2$ sheet was calculated 
using the climbing image method within the nudged elastic band (NEB) formalism 
\cite{neb}.
Diffusion of Fe adatoms on the 2D sheet at 300K was studied by means of 
Born-Oppenheimer molecular dynamics (BOMD). The temperature was adjusted
via a Nos\'{e} thermostat~\cite{nose1,nose2,nose3}. For BOMD calculations, 300 
eV was the cutoff for the plane wave basis set.
For structure optimizations and BOMD calculations, a (9 $\times$ 5 $\times$ 1) 
Monkhorst-Pack k-point mesh was used for the Brillouin zone integrations.
For NEB calculations we used a 5 $\times$ 3 $\times$ 1 k-point mesh. 

The adsorption energy $E_a$ of an Fe adatom on the sheet is defined as 
follows.
\begin{equation*}
E_a = [E({\rm Fe}) + E({\rm sheet})] - E({\rm Fe+sheet}) 
\end{equation*} 
where $E({\rm Fe+sheet})$ is the total energy of the Fe adatom adsorbed 
on the h-BNC$_2$ sheet, $E({\rm sheet})$ is the total energy of the h-BNC$_2$ 
sheet,
and $E({\rm Fe})$ is the total energy of an isolated Fe atom kept in a big box.

\section{Results and discussion}

\begin{figure}
\begin{center}
\includegraphics[scale=0.6]{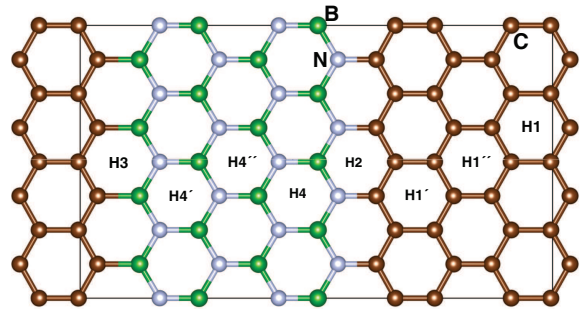}
\end{center}
\caption{(Color online) Hexagonal boron-nitride graphene (h-BNC$_2$) 
interface sheet showing different symmetry sites considered for calculation of 
diffusion barrier. Black solid line represents the unit cell considered for the calculation. 
Black (brown in color), dark grey (green in color) and white balls represent carbon, boron and nitrogen atoms respectively.  See text for more explanation.}
\label{figure:allgeom}
\end{figure}

It has been discussed in detail in the literature that the electronic structures of h-BNC$_2$ sheets, including their band gaps, depend on the 
exact arrangement of the atoms on the hexagonal lattice~\cite{bnc-ci,Zhu-11,Pooja-11}. 
In fact, it has been shown quite clearly that energetically the most favorable 
arrangement is the one in which the graphene and the h-BN regions completely phase separate~\cite{bnc-ci,Yuge-09,Zhu-11,Martins-11,Pooja-11}. 
In the chemical vapor deposition (CVD) method to produce h-BNC films, domains of h-BN were found
inside a graphene sheet~\cite{bnc-ci}. This is probably a metastable state obtained because of the kinetics of the growth process. 
It is difficult to treat such systems in a
first-principles approach, as one would require to work with large supercells. Instead, in most cases, we consider a 64-atom h-BNC$_2$ sheet in its
lowest energy configuration: the h-BN and graphene regions forming two separate domains, and being joined by zigzag C-N and C-B interfaces.
It should also be recognized that because of the periodic boundary conditions inherent in these calculations, we effectively have a 
two-dimensional heterostructure of graphene and
h-BN, the electronic properties, especially the band gaps, of which depend on the widths of these two regions~\cite{Pooja-11}. In some cases we have 
taken a sheet with two domains of graphene and h-BN each for reasons discussed later. We have also varied
the width of the h-BN region, keeping the width of the graphene region fixed, in order to study the dependence of the exchange interaction between two Fe 
adatoms on the size of the h-BN domains. The structure of a 64-atom h-BNC$_2$ sheet in its lowest energy configuration, and various high symmetry adsorption
sites on it are shown in fig. ~\ref{figure:allgeom}.

\subsection{Adsorption and diffusion of individual Fe adatoms}

In an earlier work~\cite{Pooja-11} we reported the adsorption energies of a single Fe adatom at various symmetry sites on h-BNC$_2$ sheets.
On a 32-atom supercell, a hexagonal site within the graphene region turned out to be the most favorable. 
However, on larger supercells, the hexagonal site at the C-B interface was found
to be most favorable. On these larger supercells there are many more symmetry inequivalent hexagonal sites at which an adatom may be adsorbed. In this work we calculated adsorption energies of a single Fe adatom at all the inequivalent hexagonal
sites on a 64-atom supercell (fig. ~\ref{figure:allgeom}). Adsorption energies at these sites, and the heights of the Fe adatom above the BNC sheet are reported in table ~\ref{table:bind-Fe}.
We would like to clarify here that although we started our calculations by placing the Fe adatom at the center of the hexagonal sites marked in fig.~\ref{figure:allgeom}, after
relaxation the adatom goes off-center in some cases. Subsequent calculations of diffusion barriers by the NEB method are carried out with these relaxed
positions of the Fe adatom as the initial and final positions. Among the hexagonal sites, H4$'$ turns out to be an unstable one.  It is seen from our calculations that the H3 site at the C-B interface is much lower in energy (deeper potential) than H4$'$ and hence prevents the formation of an energy barrier from going to H4$'$ to H3. Thus an adatom placed at this site moves to the neighboring H3 site at the C-B interface after relaxation. Our calculations clearly show that the H3 site at the C-B interface is the most favorable one followed closely by the H1 site. As found earlier on a 32-atom supercell, the H2 and H4 sites are much less favorable than H1 and H3. In fact, the adsorption energy at all the hexagonal sites within the h-BN
region is rather small. 

\begin{table}
\caption{Adsorption energy of Fe adatom ($E_a$) at different sites. The most stable site is marked in bold. See text for details.}
\label{table:bind-Fe}
\begin{tabular}{|l|c|c|}
\hline\hline
Site & $E_a$ (eV) & Height (\AA)\\
\hline
H1 & 1.65 & 1.57 \\
H1$'$ & 1.36 & 1.58 \\
H1$''$ & 1.37 & 1.55 \\
H2 & 0.89 & 1.65 \\
{\bf H3} & {\bf 1.74} & {\bf 1.72} \\
H4 & 0.39 & 1.62 \\
H4$''$ & 0.57 & 1.90\\
\hline
\end{tabular}
\end{table}

We calculated barriers between all the nearest neighbor hexagonal sites on the 
h-BNC$_2$ sheet as shown in fig.~\ref{figure:allgeom} in both directions using the NEB 
method. For NEB calculations we have taken straight line path along the line 
joining two hexagonal symmetry sites. 
The calculated barrier values are given in table ~\ref{table:barrier}. It is found that all the energy barriers are rather small. The largest barrier is for movement from H1$'$ to H1$''$ 
within the graphene region with a value of 1.3 eV. The H1$'$ - H1$'$ barrier is also similar and is 1.08 eV. H1$'$ - H2 diffusion barrier is 0.53 eV, 
but in contrast the H2 - H1$'$ barrier is only 0.1 eV. It seems that it is easier for an adatom to reach the H1$'$ site than to move out of it. Table ~\ref{table:barrier} suggests that diffusion barriers for both reaching and leaving H3 site are moderate, and one may
not expect this site to play any special role in finite temperature diffusion processes. Another feature that comes out of the results presented in table ~\ref{table:barrier} is that
the diffusion barriers in the h-BN region are generally smaller than those in the graphene region. This obviously suggests that the Fe adatoms will be more
mobile in the former region.

\begin{table}
\caption{Calculated diffusion barriers (in eV) along different paths on the h-BNC$_2$ sheet.} 
\label{table:barrier}
\begin{tabular}{|c|c||c|c||c|c|}
\hline\hline
Path & Barrier & Path & Barrier & Path & Barrier\\
\hline
H3 -- H1 & 0.34 & H1 -- H3 & 0.26 & H4$''$ -- H4$''$ &  0.21 \\
H2 -- H1$'$ & 0.1  & H1$'$ -- H2 & 0.53 & H3 -- H3 & 0.6 \\
H1 -- H1$''$ & 0.75  & H1$''$ -- H1 & 0.53 & H2 -- H2 & 0.13 \\
H1 -- H1 & 0.32 & H1$'$ -- H1$'$ & 1.08 & H1$'$ -- H1$''$ & 1.3 \\
 H1$''$ -- H1$''$ & 0.46 & H4 -- H4 & 0.17  &   &
 \\
\hline
\end{tabular}
\end{table}

Our NEB calculations at zero temperature suggest that a single Fe atom is likely to be trapped at  the H1$'$ site. It has been argued in the literature that the exact
mechanism of diffusion at finite temperatures could actually be more complex~\cite{Zobelli-07}. 
In fact, one should calculate the difference between the Gibbs' free
energies, rather than the internal energies, of the local minima and the saddle point to estimate the diffusion barrier at finite temperatures. 
The diffusion barrier may increase
or decrease, or even show non-monotonic behavior with increasing temperature depending on the details of the diffusion mechanism and the
vibrational modes at the saddle points in a given system. Rather than estimating diffusion barrier of the adatoms, we have studied their
real-time dynamics on the sheet at finite temperatures through MD. Since most practical applications will be at ambient conditions,
we set the temperature to $T=300$ K. We started our BOMD simulations by placing an Fe adatom at all the inequivalent hexagonal sites 
at an initial time $t=0$. 
When we started the simulations with the Fe adatom at a H1$'$ or a H1$''$ site, the
adatom remains stuck there. At least up to 15 ps, the Fe adatom could not cross the barrier and move out of these sites. 
If the Fe atom is placed at a H2 site, it diffuses to H1$'$ and gets trapped. 
Starting from the H1 site, the Fe adatom moves back and forth between H1 and H3 before finally settling at a H3 site. 
For all other sites (mainly hexagonal sites in BN part)  the Fe adatom diffuses 
rapidly and eventually gets trapped at the H3 site at the C-B interface.
In particular, as our calculated barriers suggest, the Fe adatom can diffuse rather easily
over the h-BN region and can reach the C-B interface. The observation that it gets stuck at an H3 site is consistent with the fact that 
the H3 site has the largest adsorption energy. However, how a Fe adatom crosses the relatively larger barriers at the H1$'$ site in its motion from 
the h-BN region across the graphene region to a H3  site is intriguing. It is likely
that the zero temperature NEB calculations fail to capture the full complexity of the motion of all the atoms at finite temperature. In light of the
discussion in Ref.~\cite{Zobelli-07}, the Gibbs' free energy difference between the local minimum and the saddle point is smaller at T=300 K
compared to that at T=0 K. Another way of looking at the process is that a concerted motion of all the atoms at finite temperatures, particularly when a Fe adatom approaches a H1$'$ site from the h-BN direction, reduces the effective diffusion 
barriers to an extent that these are crossed at room temperature. 
 
It is quite interesting to note that a single Fe 
adatom gets adsorbed at the H3 site, found to be the most favorable 
energetically, even at room temperature. Given this scenario there are two 
pertinent questions one needs to address. First, the nature of magnetic coupling 
between two Fe adatoms adsorbed at two different
H3 sites along the C-B interface. This will give a preliminary insight into the possible magnetic self assemblies one may obtain on this sheet. Second, whether
two such Fe adatoms remain stuck at two different H3 sites at finite temperatures, or whether they diffuse along H3 - H3 path parallel to the C-B interface and form a dimer.

To answer the first question we put two Fe adatoms at all possible combinations of H3 sites. In particular, in a 64-atom supercell as shown in fig.~\ref{figure:allgeom}, we put
them at the nearest neighbor (NN) and next nearest neighbor (NNN) H3 sites. For each of these possibilities we calculated energies for the parallel (FM) and antiparallel (AFM) alignments of the spins on the Fe adatoms. 
It turns out that a FM alignment of the spins when two Fe adatoms occupy NN H3 sites has the lowest energy. Each Fe adatom
has a moment of 3 $\mu_B$ in this case. An antiparallel alignment of spins with 
the Fe atoms at NN sites is $\sim$0.53 eV higher, which indicates to a very 
large nearest neighbor ferromagnetic coupling. In this case, the two
Fe adatoms are at slightly different heights from the sheet, and the moments on the two Fe atoms turn
out to be different: 3.15 $\mu_B$ and 2.64 $\mu_B$. All other arrangements of the atoms are $\sim$2 eV higher in energy. As for the second question,
we studied the diffusion of two Fe adatoms placed at two NNN H3 sites at a C-B interface. 
Total magnetic moment in this case is 4.0 $\mu_B$ with individual Fe adatom moments being 2.43 $\mu_B$ and 2.6 $\mu_B$. 
Up to 30 ps, the adatoms remain stuck at their respective H3 sites, and only execute thermal oscillations around mean positions. This is also evident from a high value of diffusion barrier (0.6 eV) between two H3 sites as reported in table ~\ref{table:barrier}. Thus the H3 sites truly act as trapping sites for individual Fe adatoms diffusing on the h-BNC$_2$ sheet at room temperature.

\begin{figure}[ht]
\begin{center}
\includegraphics[scale=.5]{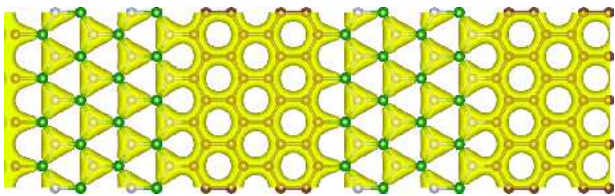}
\end{center}
\caption{(Color online) Charge density of pure graphene / h-BN sheet.}
\label{figure:charge}
\end{figure}

In order to investigate why C-B interface is more favorable than C-N, we have plotted the charge density profile for the pure graphene / h-BN sheet (\ref{figure:charge}). It is clearly seen that at the C-B interface, the charge density is not homogeneously distributed over the C-B bond whereas at the C-N interface, one finds a homogeneous charge distribution ($\pi$ character originating from C and N $p_z$ orbitals) over the C-N bond. The asymmetry in the distribution of charge density allows C-B interface to be more reactive. 

\begin{figure}[ht]
\begin{center}
\includegraphics[scale=0.32]{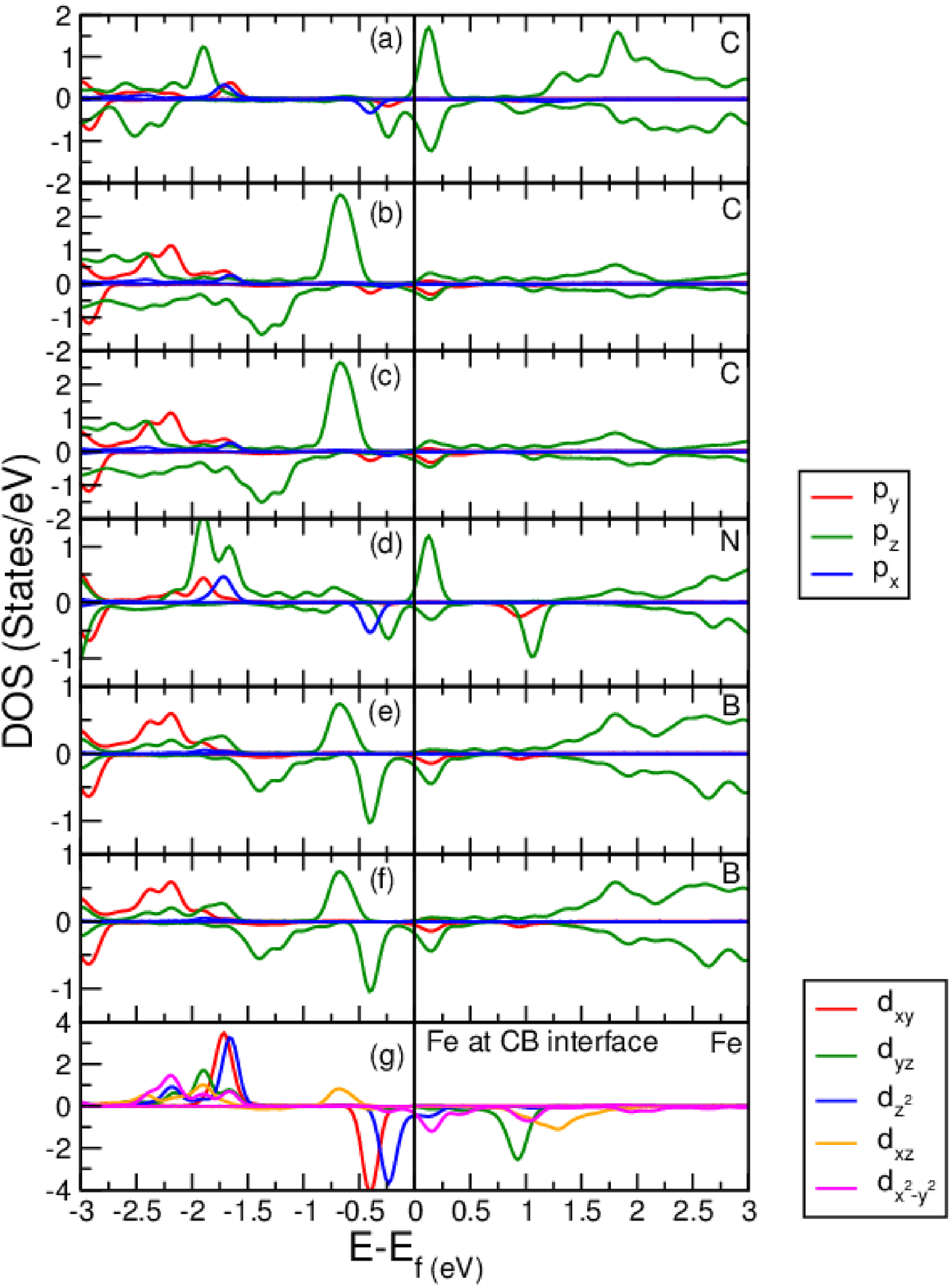}\\
\includegraphics[scale=0.32]{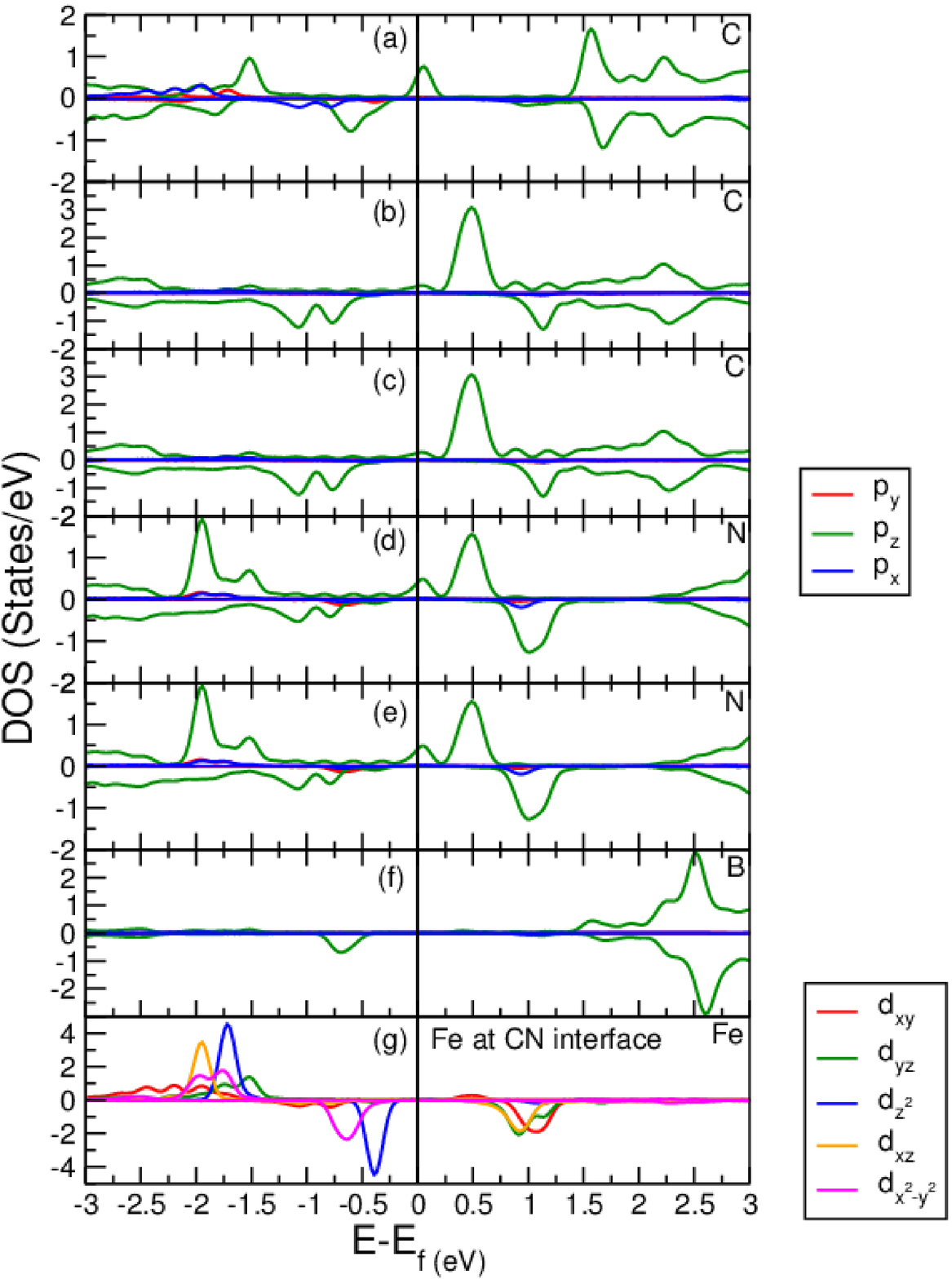}\\
\end{center}
\caption{(Color online) m$_l$ projected DOS for Fe at C-B and C-N interface along with the same for C, B and N atoms belonging to the respective hexagon, whose center is occupied by Fe.}
\label{figure:mldos}
\end{figure}

A close inspection with the help of m$_l$ projected DOS for adsorption of an Fe atom separately
at the C-B and C-N interfaces reveals interesting features (\ref{figure:mldos}). For the C-B interface, Fe $d_{xy}$ orbital hybridizes with B p$_z$ orbital quite prominently. Also, Fe $d_{z^2}$ orbital hybridizes with C $p_z$ and N $p_z$ orbitals. So, a relatively strong binding scenario is established. For the C-N interface,
no significant hybridization (except a small one between Fe $d_{x^2-y^2}$ and B and C p$_z$ orbitals)
between Fe $d$ orbitals and $p$ orbitals of B/C/N is observed. 

\subsection{Diffusion of multiple Fe adatoms}

\begin{figure}[ht]
\begin{center}
\includegraphics[scale=0.22]{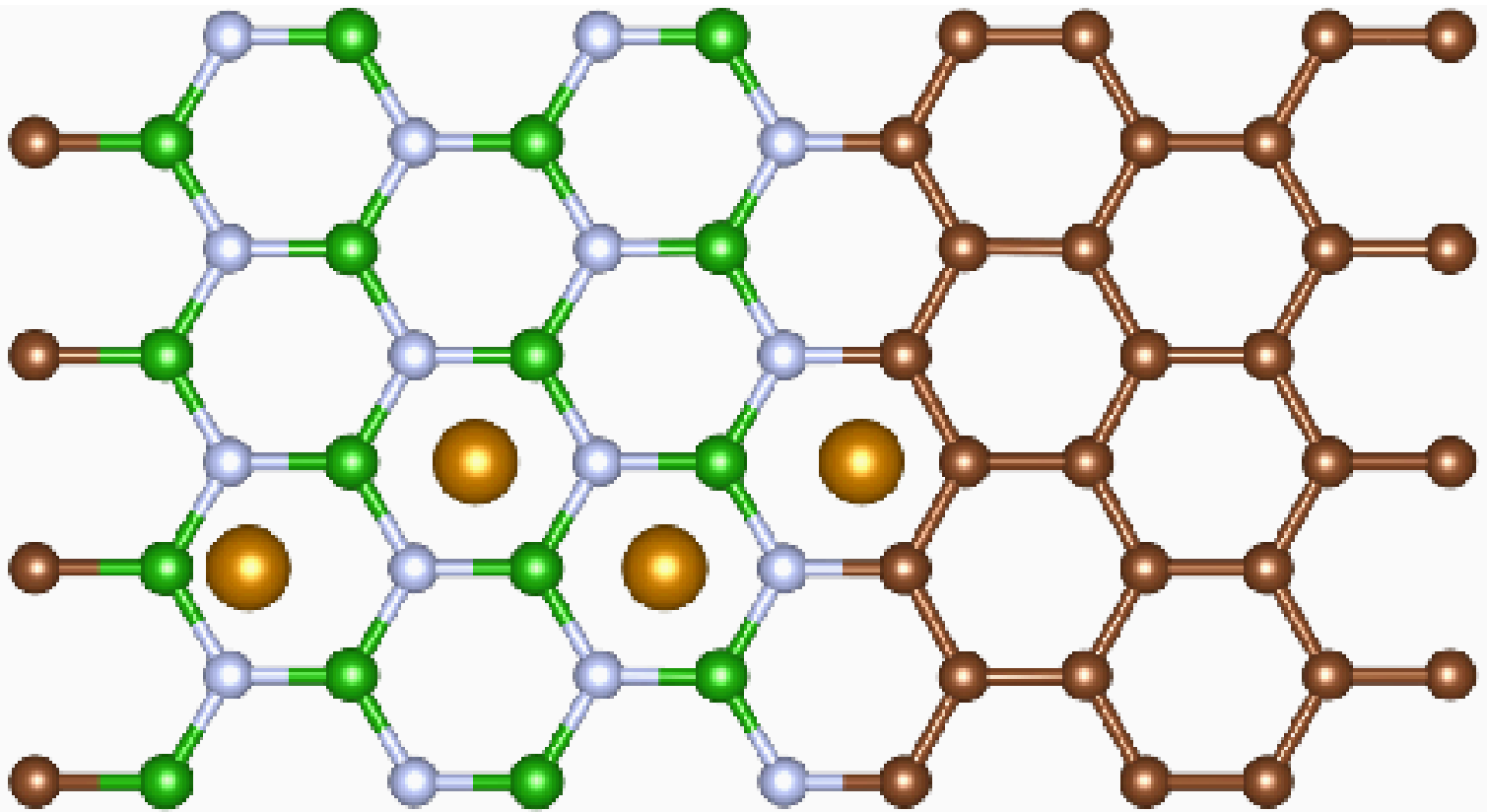}
\includegraphics[scale=0.11]{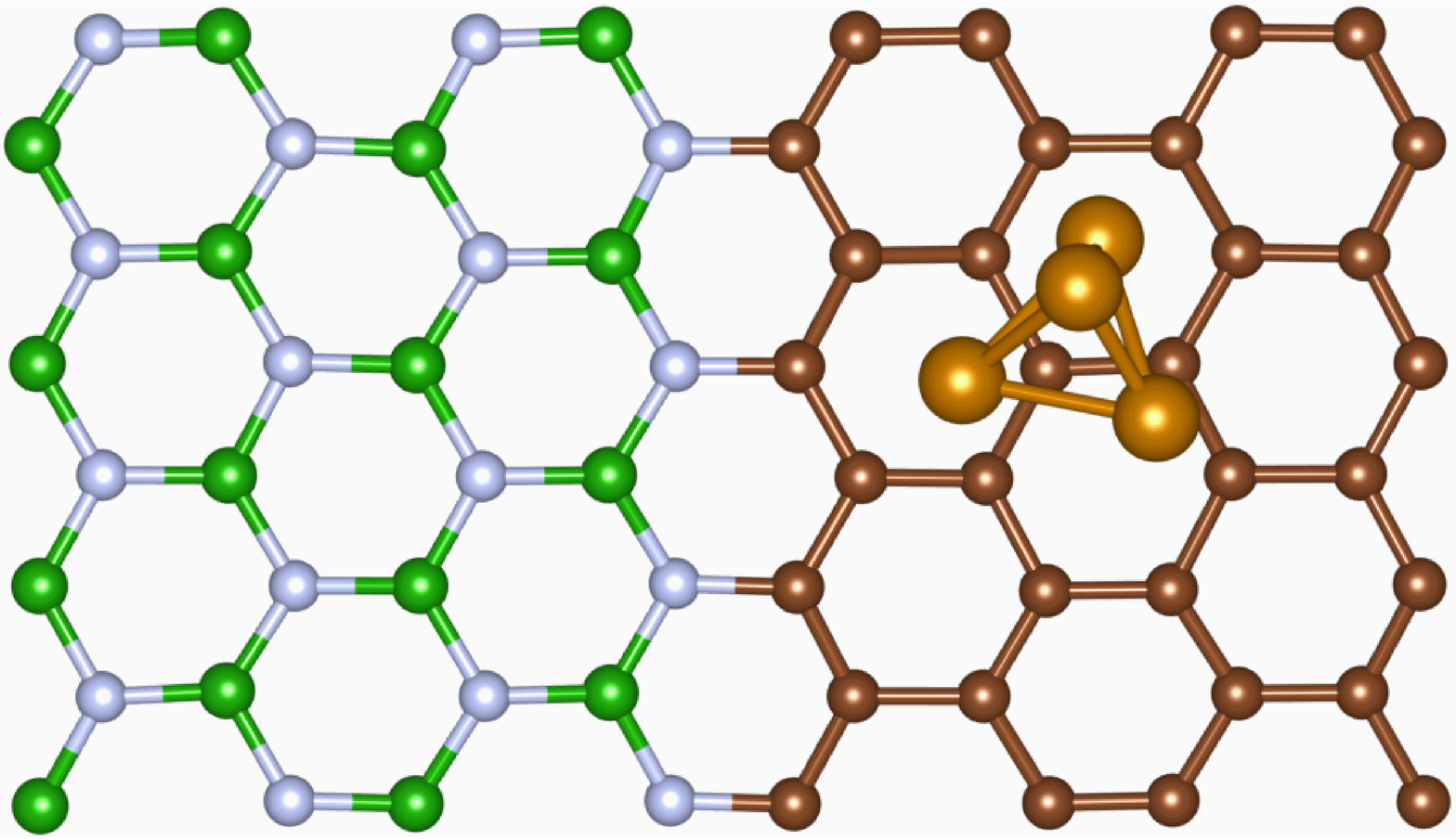}\\
(0 ps) ~~ (3 ps)\\
\includegraphics[scale=0.11]{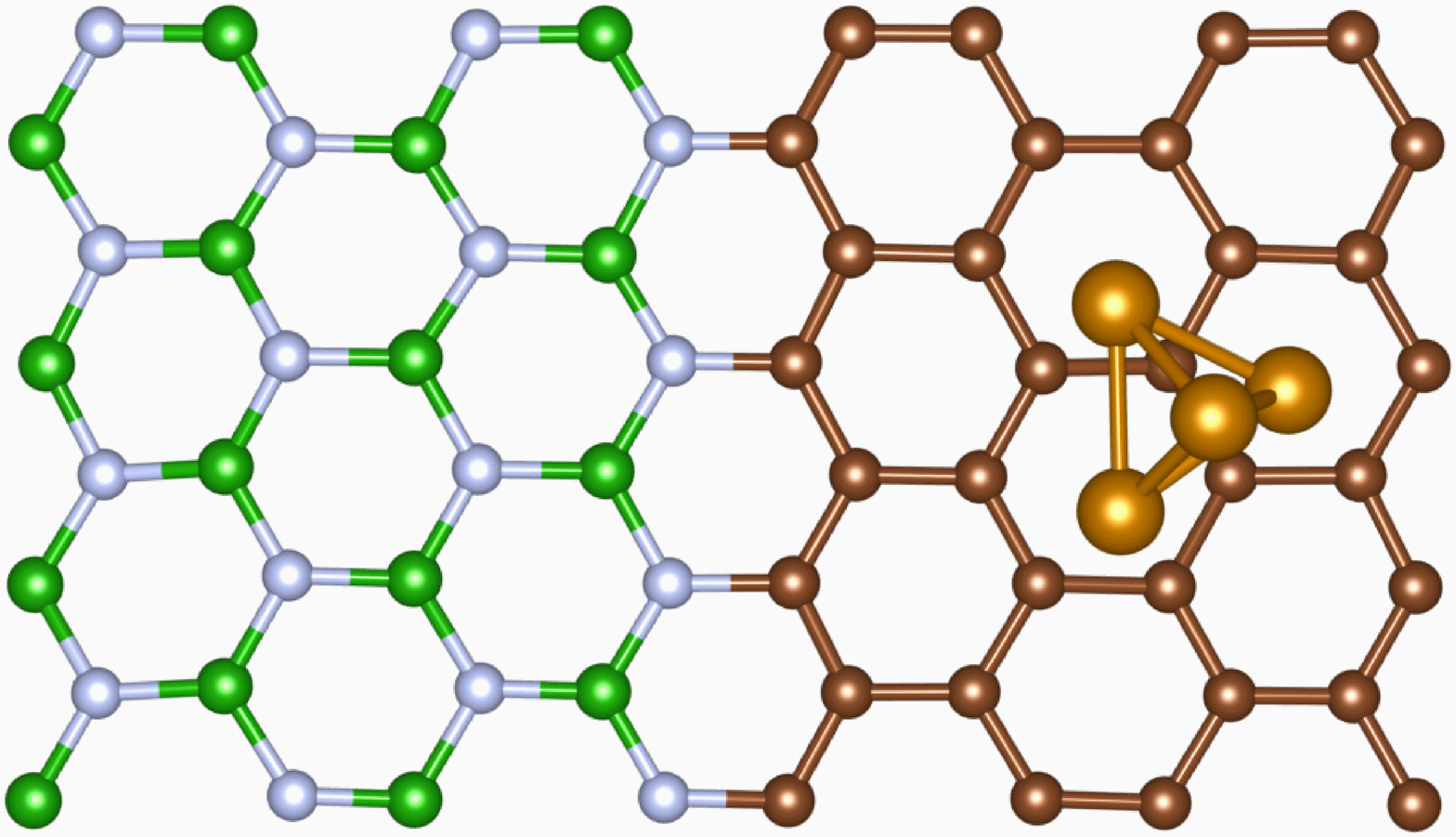}
\includegraphics[scale=0.22]{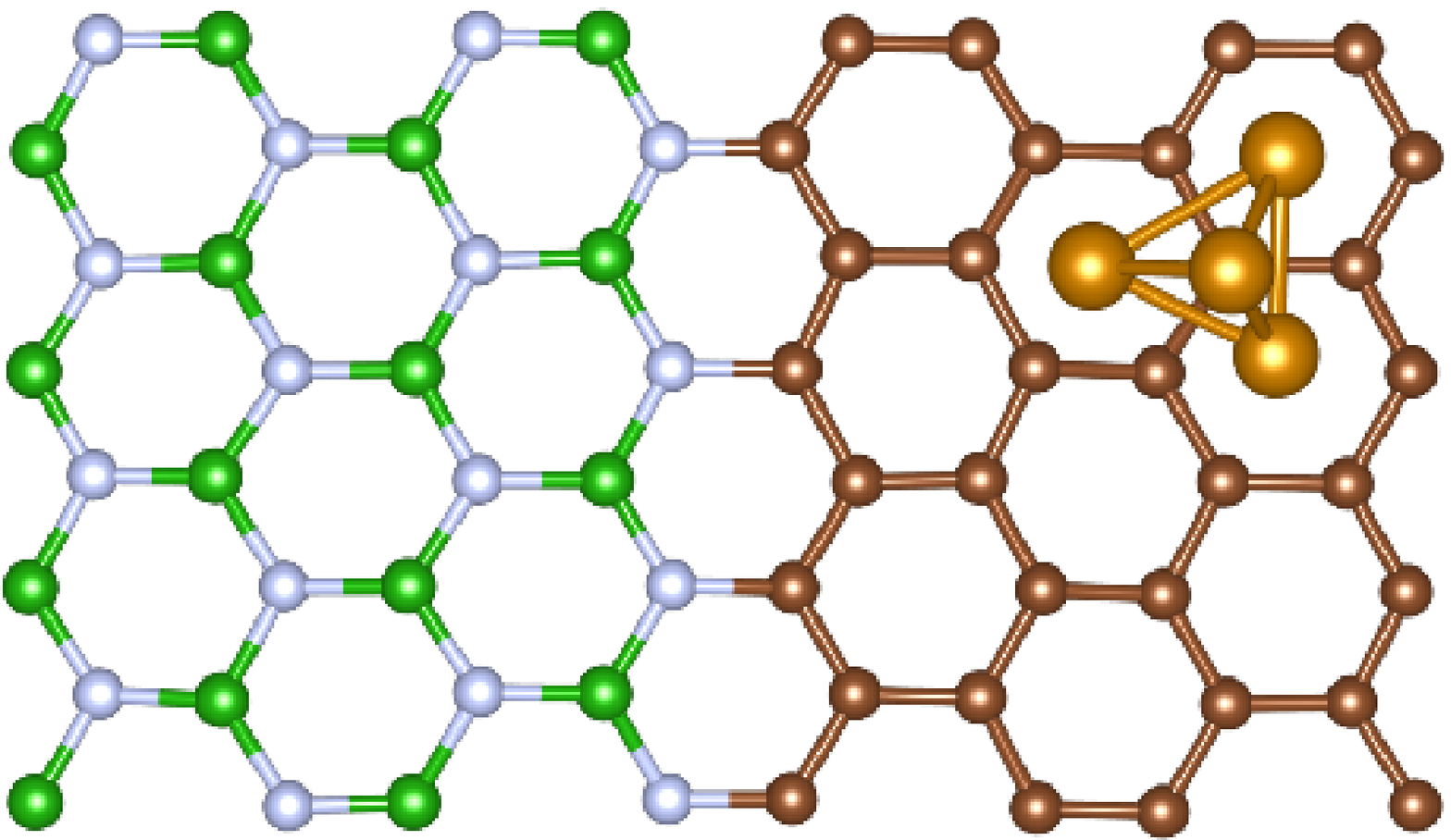}\\
(9 ps) ~~ (15 ps)\\
\end{center}
\caption{(Color online) Time evolution of 4 Fe atoms on BNC sheet starting from the BN part. }
\label{figure:time-evo-bn}
\end{figure}

Having gained insights into the finite temperature dynamics of isolated Fe adatoms, and magnetic exchange coupling between two isolated 
Fe adatoms on a h-BNC$_2$ sheet, we now try to address the crucial question of clustering and subsequent dynamics when 
more than one Fe adatoms are deposited
on the sheet. Particularly striking was our finding that the H1$'$ site, in spite of having large barriers, do not trap Fe adatoms, which ultimately get stuck at the H3 sites. This makes finite temperature BOMD calculations for a number of Fe adatoms even more relevant. 
\begin{figure}[ht]
\begin{center}
\includegraphics[scale=0.22]{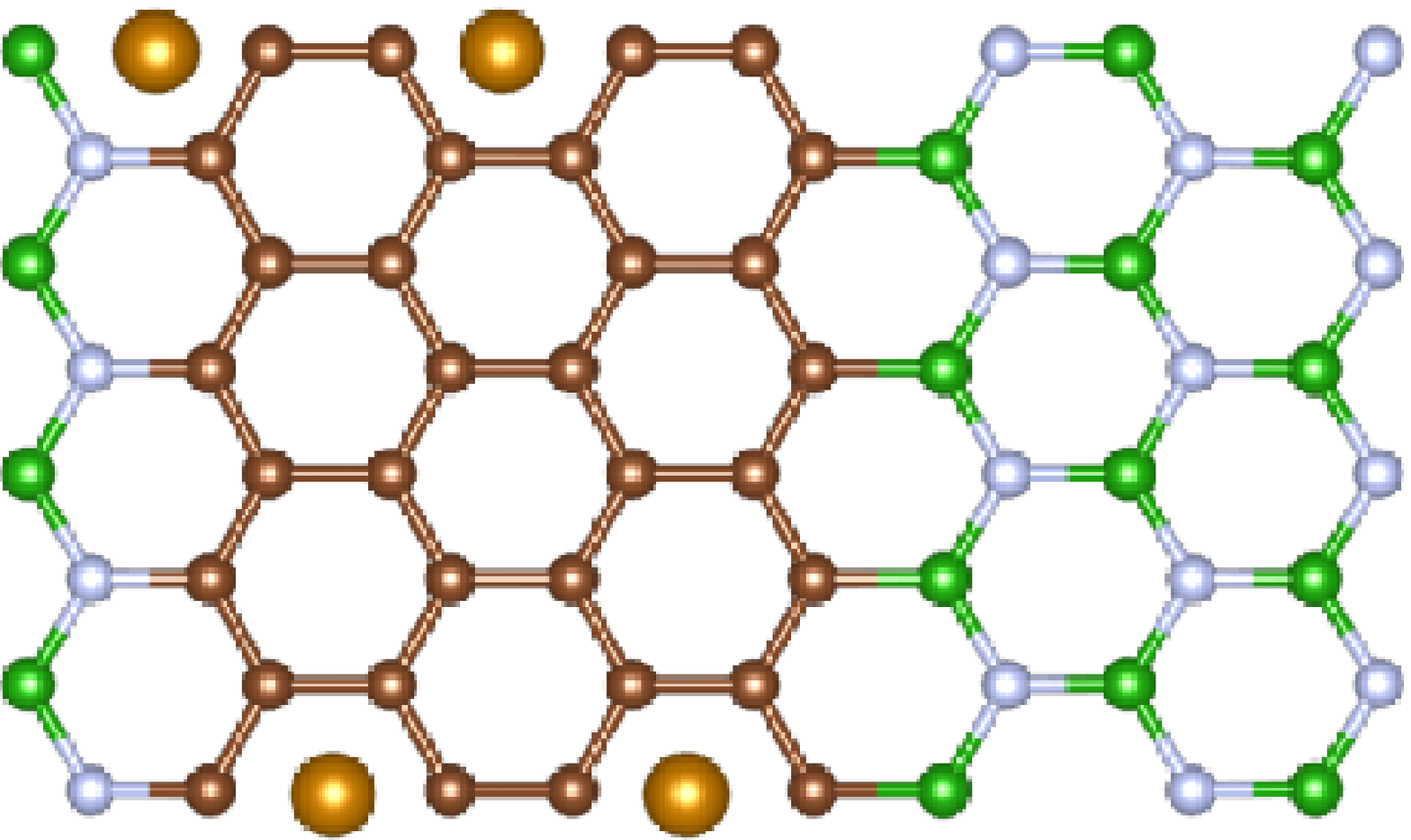}
\includegraphics[scale=0.22]{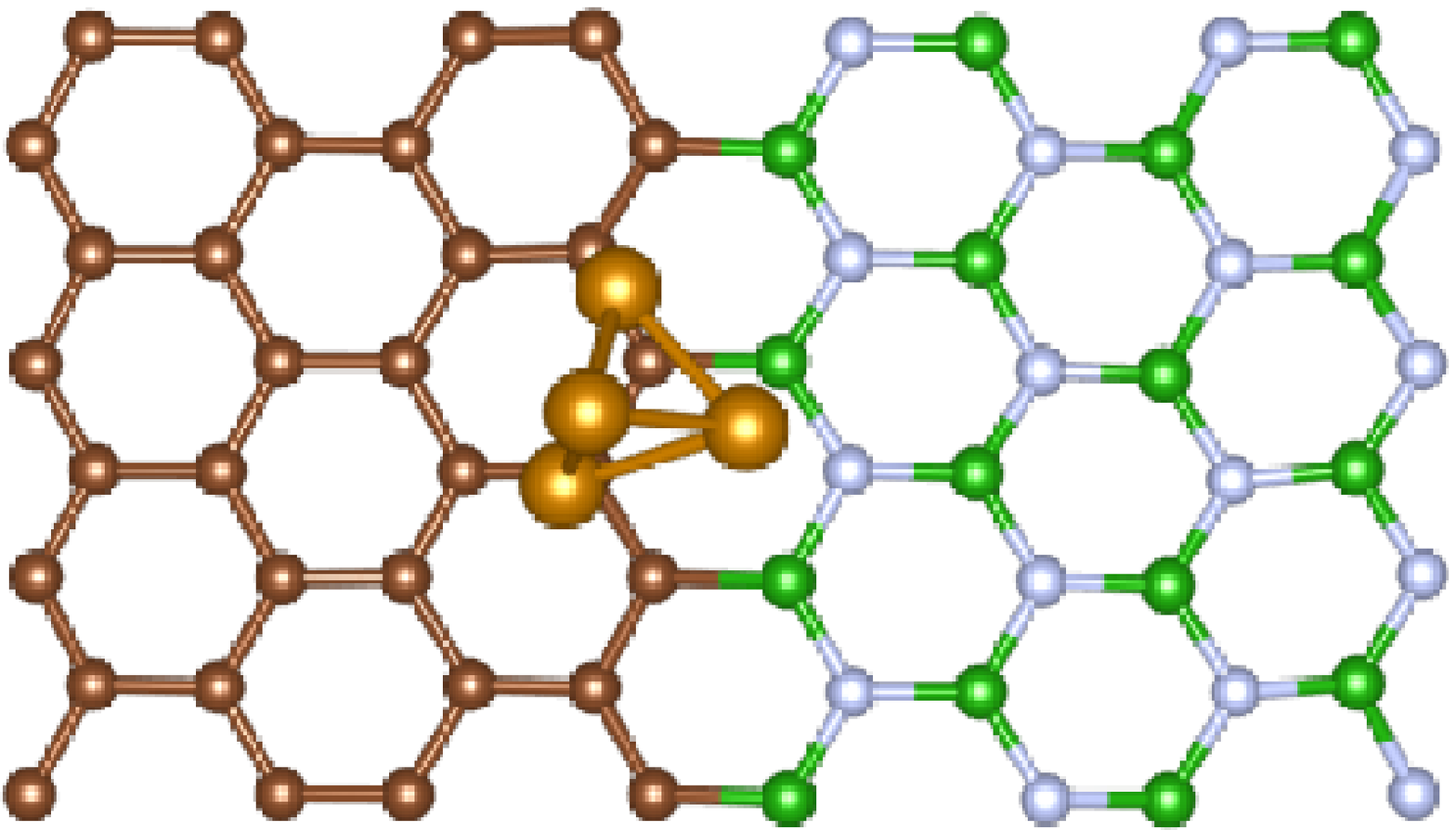}\\
(0 ps) ~~ (3 ps)\\
\includegraphics[scale=0.22]{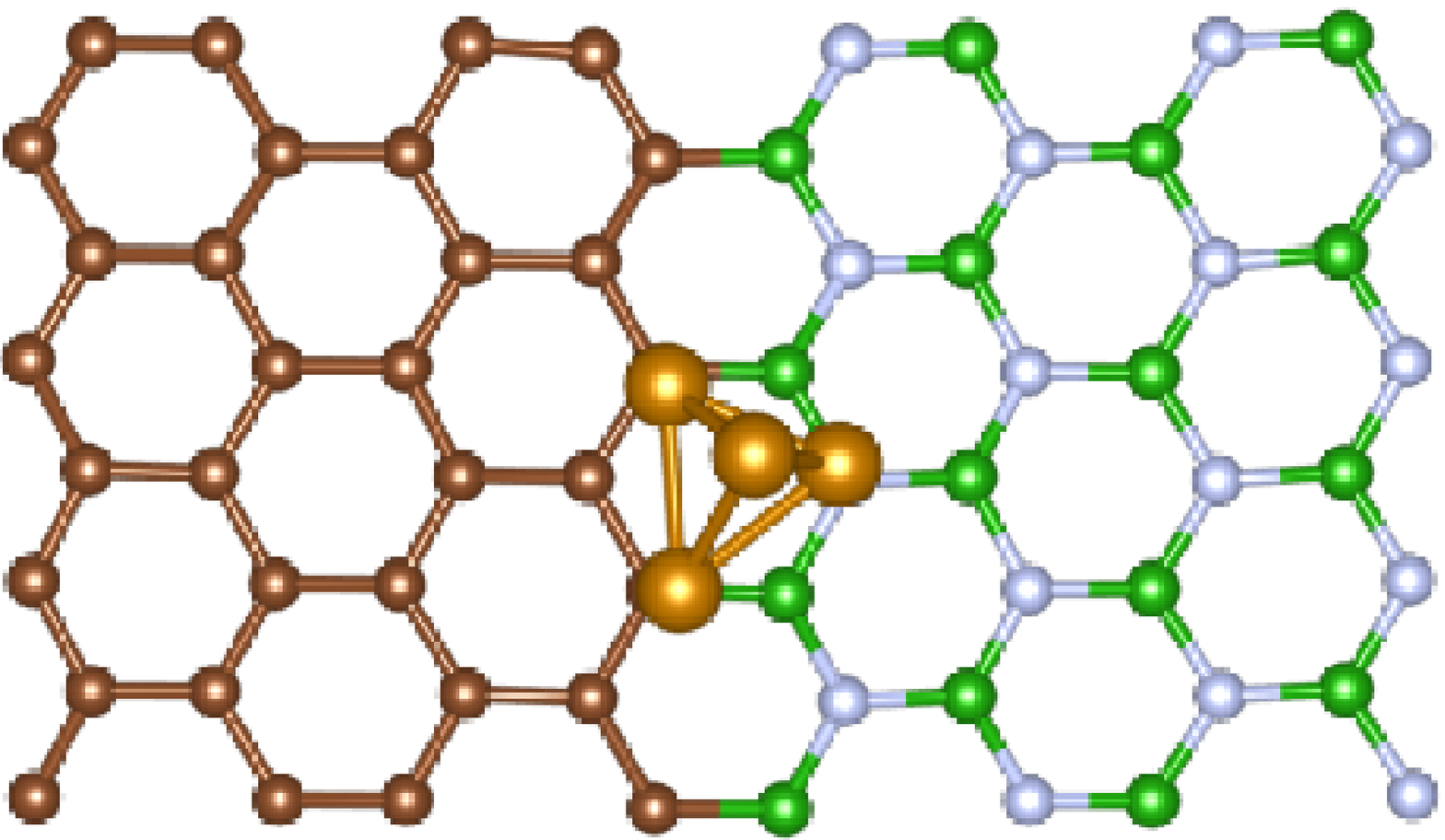}
\includegraphics[scale=0.22]{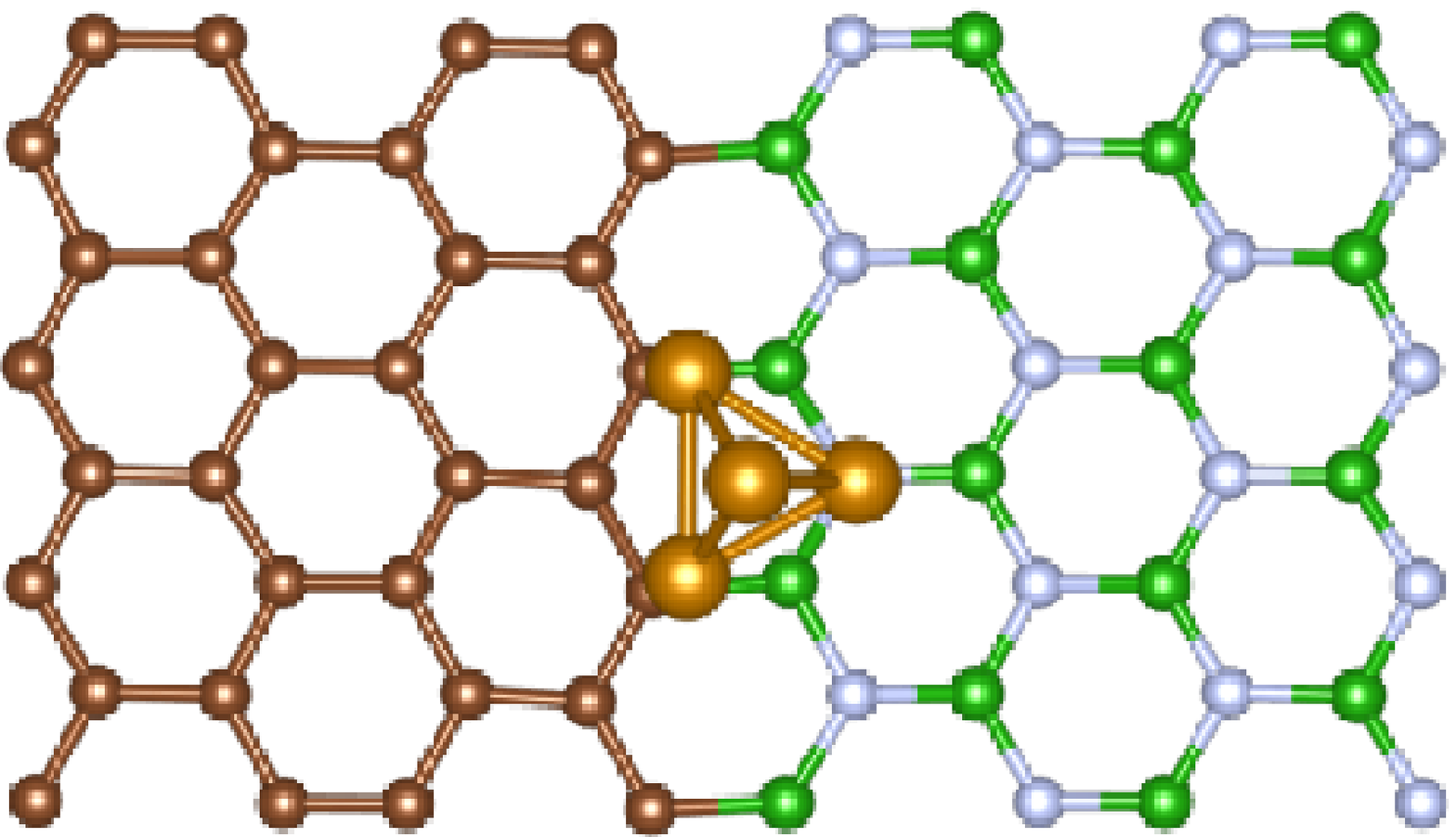}\\
(9 ps) ~~ (15 ps)\\
\end{center}
\caption{(Color online) Time evolution of 4 Fe atoms on BNC sheet starting from the graphene part.}
\label{figure:time-evo-gr}
\end{figure}

A summary of these calculations are presented 
in fig.~\ref{figure:time-evo-bn} and fig.~\ref{figure:time-evo-gr}. In one case we placed four Fe adatoms at four different hexagonal sites in the h-BN region as shown in the first
panel in fig.~\ref{figure:time-evo-bn}. After we start the BOMD calculations, the adatoms diffuse rather fast, and  roughly after 2.1 ps all the four Fe adatoms come together to form a cluster and possess a total magnetic moment of 11.4 $\mu_B$. Once the cluster is formed, it moves as
a single unit across the h-BN and graphene regions towards the C-B interface. MD snapshots after 3 ps, 6 ps (not shown) and 9 ps show this clearly. After about 12 ps
the cluster reaches the C-B interface and subsequently it only executes thermal oscillations around its mean position up to 15 ps. In another calculation we put four Fe adatoms at four hexagonal sites in the graphene region as shown in fig.~\ref{figure:time-evo-gr}.
Interestingly, in this case also they diffused very rapidly and formed a Fe$_4$ cluster roughly after 1.5 ps. Since the adatoms were initially placed in the
graphene region, the Fe$_4$ cluster reached the C-B interface rather quickly, within 3 ps, and did not move subsequently at least up to the time we continued
the BOMD calculations, {\it i.e.}, 15 ps.
\begin{figure*}[ht]
\begin{center}
\includegraphics[scale=0.48]{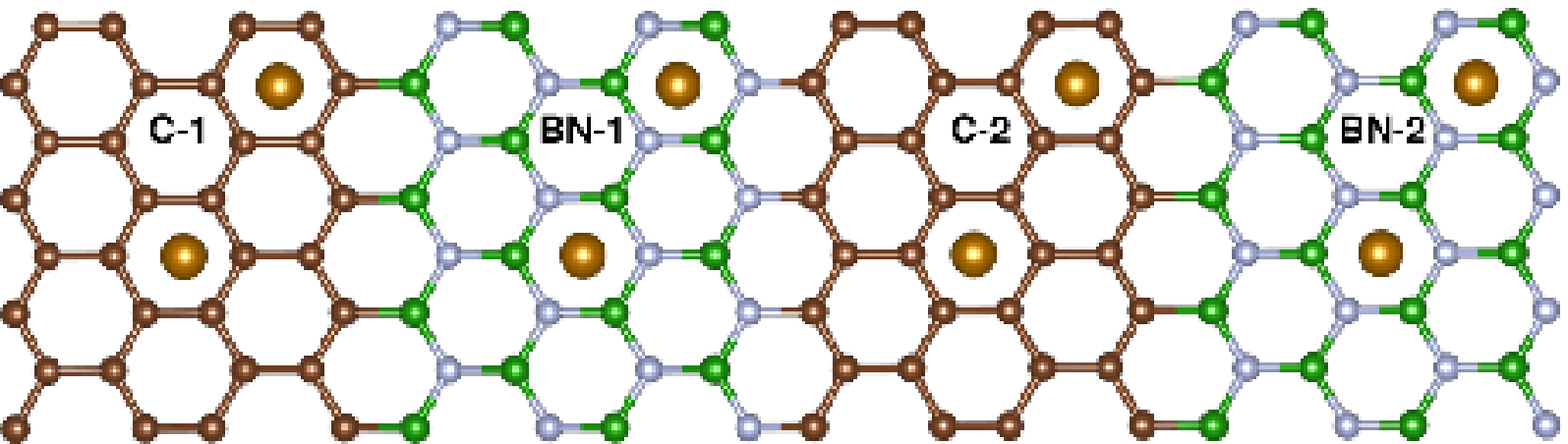}  \hspace*{0.2in}
\includegraphics[scale=0.48]{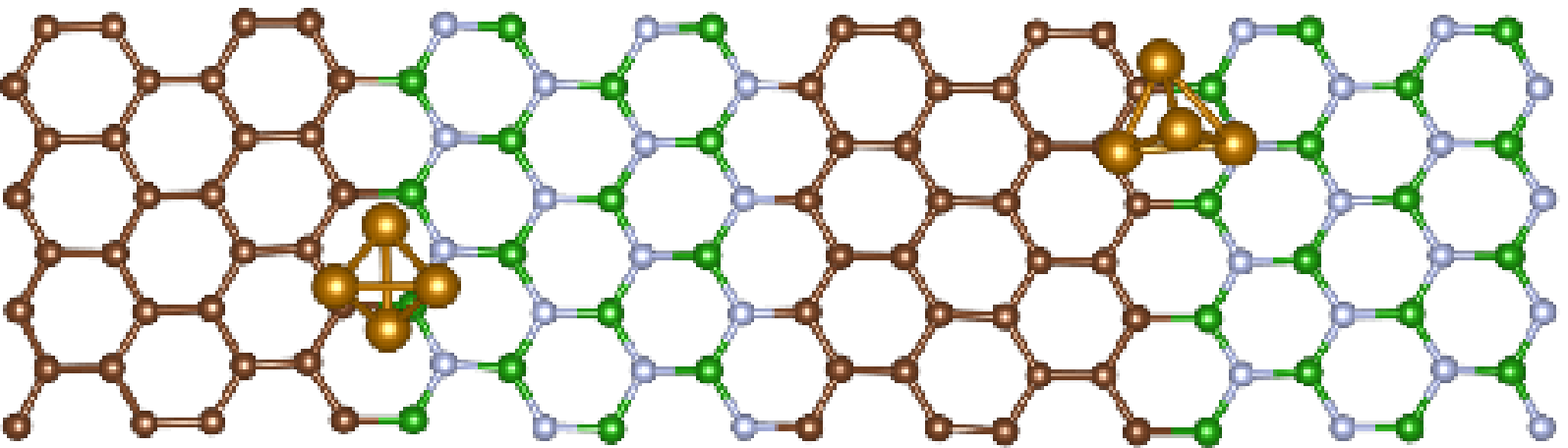}\\
 (0 ps) ~~ (3 ps) \\
\includegraphics[scale=0.48]{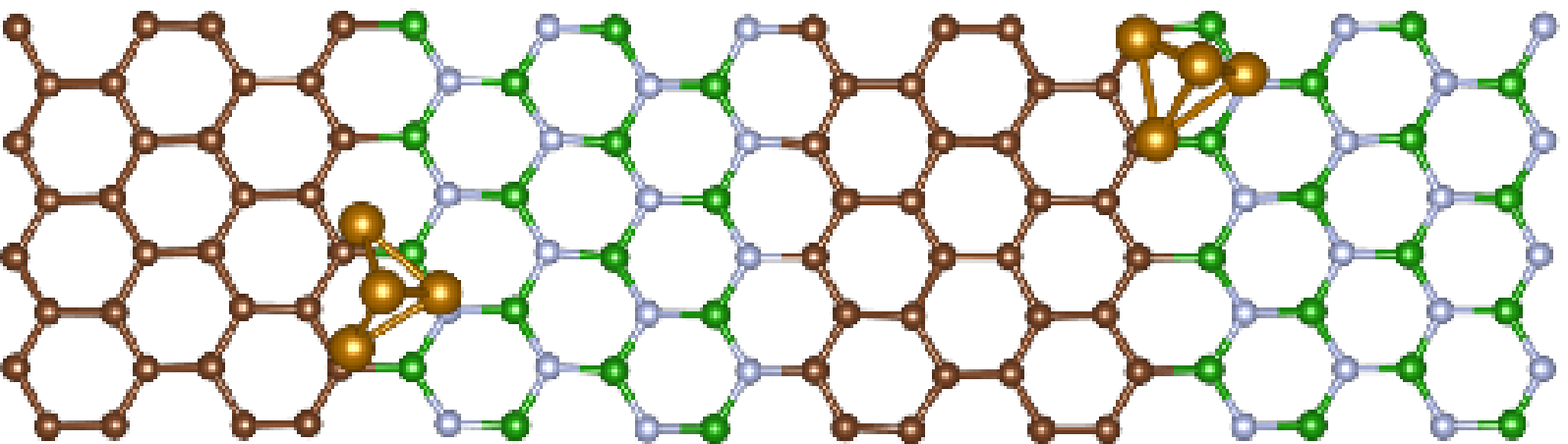}   \hspace*{0.2in}
\includegraphics[scale=0.48]{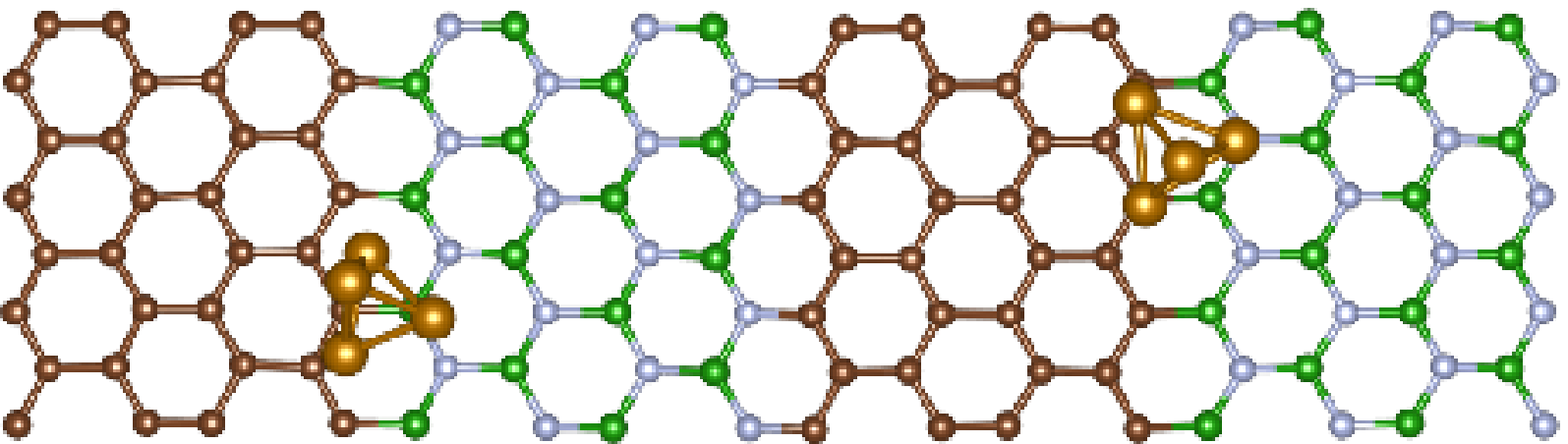}\\
(9 ps) ~~ (15 ps)\\
\end{center}
\caption{(Color online) Time evolution of 8 Fe atoms on the BNC sheet. We placed 2 Fe 
atoms separated from each other on 4 patches of BN and Graphene. See text for discussions.}
\label{figure:time-evo-broad}
\end{figure*}

Our BOMD calculations confirm the earlier results that Fe adatoms on a h-BNC$_2$ sheet prefer to cluster together. 
The interesting new insights these calculations 
provided are how quickly the cluster is formed, and the fact the cluster as a whole diffuses easily and finally gets stuck at the C-B interface. 
In the above calculations
all four Fe adatoms were placed either in the h-BN or graphene regions. When Fe adatoms are deposited on a h-BNC$_2$ sheet in experiments, one will not have such
microscopic control over where they actually land. Moreover, in view of the fact that CVD experiments produce domains of h-BN inside a graphene sheet, the adatoms
are equally likely to land on either of the regions. To simulate such possibilities in our calculations we consider a larger sheet with 128 atoms: 64 C, 32 B and 32 N atoms.
These are divided into two graphene and two h-BN regions as shown in fig.~\ref{figure:time-evo-broad}. Eight Fe adatoms are placed at eight 
hexagonal sites on this sheet at
$t=0$ in the following way. Two Fe adatoms are placed in each of the C-1 and C-2 
regions, and two Fe adatoms each in the BN-1 and BN-2 regions.

Note that there is a C-B interface between C-1 and BN-1 regions (henceforth 
referred as C-B-int1), a C-N interface between BN-1 and 
C-2 regions (henceforth referred as C-N-int1), another C-B interface between 
C-2 and BN-2 (C-B-int2) and finally another C-N interface 
between BN-2 and C-1 (C-N-int2) due to the periodic 
boundary conditions. As seen in the second panel of fig.~\ref{figure:time-evo-broad}, 
within 3 ps, all the four Fe adatoms in the BN-1 and C-1 regions form a Fe$_4$ 
cluster. The two Fe adatoms placed in the BN-1 region easily move towards 
the C-B-int1. At the same time the two Fe adatoms placed in the C-1 region 
also move towards the C-B-int1. Then these four Fe atoms forms a Fe$_4$ 
cluster and as a unit gets trapped at the C-B-int1 interface. Up to 15 ps 
simulation time, the cluster can be seen to be stuck there in fig.~\ref{figure:time-evo-broad}. 
Similarly, the other four Fe adatoms placed in the BN-2 and C-2 regions form a 
second Fe$_4$ cluster near the C-B-int2 interface (two Fe adatoms coming from 
BN-2 region and other two Fe adatoms coming from C-2 region) which gets trapped 
at the C-B-int2 interface. 
These findings throw some very important light on the motion and 
behavior of Fe adatoms on the kind of h-BNC sheets produced in the CVD 
experiments. These suggest that the Fe adatoms within h-BN and graphene regions separated by 
a C-B interface will form a cluster and will 
eventually get trapped at a C-B interface.

\subsection{Exchange interaction}
\begin{figure}[ht]
\begin{center}
\includegraphics[scale=0.32]{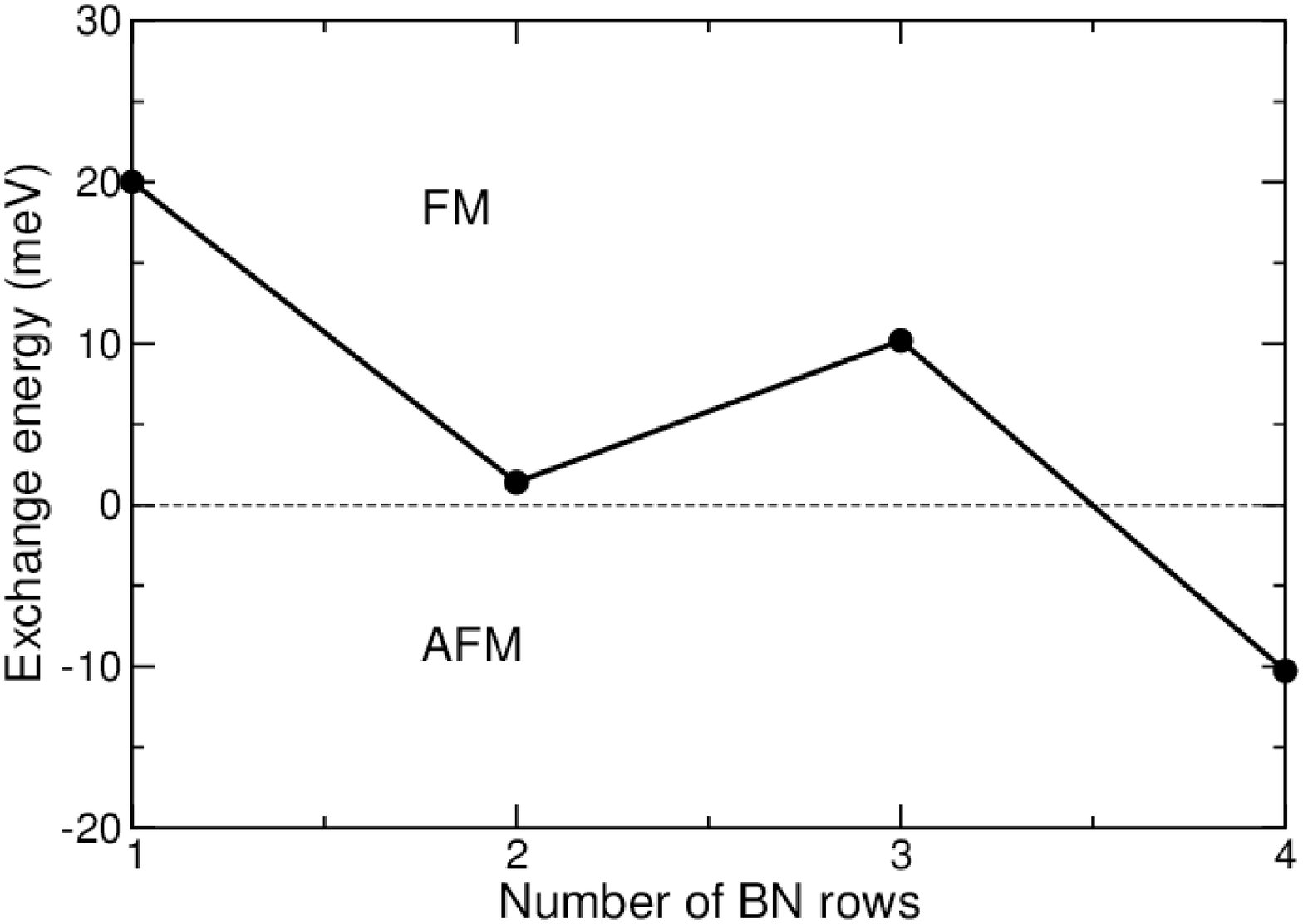}\\
\end{center}
\caption{Exchange energy between two Fe$_4$ clusters placed at C-B interfaces 
separated by h-BN domains of different widths in a h-BNC$_2$ sheet. E$^{Exch}$ 
= E$^{AFM}$ - E$^{FM}$}
\label{figure:exchange}
\end{figure}

We now discuss the next important question about the nature of magnetic interaction between different clusters trapped at different points at the same or
different C-B interfaces. This is obviously a complex issue in reality, particularly on a 2D sheet in which there are many h-BN domains of different sizes and shapes, and many
C-N and C-B interfaces. In order to gain an understanding into this important aspect we study the nature of coupling between two Fe$_4$ clusters trapped at 
two different C-B interfaces as a function of varying width of the h-BN regions (shown in fig.~\ref{figure:time-evo-broad}) in between. The strength of the exchange interaction between 
the two Fe$_4$ clusters is measured by the difference in the total energies of the AFM (E$^{AFM}$) and FM (E$^{FM}$) alignments of the moments of the clusters. Here, we have assumed that the intra-cluster exchange is ferromagnetic as observed in free clusters.  The exchange energy is
plotted as a function of the widths of the h-BN regions in fig.~\ref{figure:exchange} while the width of the graphene region is kept fixed at 4 zigzag C rows. 
It is clear that for smaller widths of the h-BN regions, the exchange is FM. At intermediate
widths, when there are two BN rows in the h-BN region, the FM and AFM energies are fairly close implying a very weak exchange coupling. For three BN rows the coupling
becomes relatively stronger and is still FM. For larger separations the exchange becomes AFM. The explanation of this observation is given below. 
At all separations between the Fe$_4$ clusters 
calculated here, however, the exchange coupling is rather small as the difference between AFM and FM energies is only a few tens of meV.  
This suggests that in a h-BNC sheet as prepared in experiments, complex magnetic arrangement of different clusters may emerge but only at low temperatures. 

At a first glance, one would expect an exponentially decaying exchange interaction between two magnetic clusters separated by a wide band gap material (h-BN) as observed in the case of dilute magnetic oxides, e.g., Co doped ZnO \cite{rmp}. However, we observe a non-monotonous behavior of exchange interaction in the present case. This is related to the peculiar electronic structure at the interface between graphene and h-BN (shown in fig.~\ref{figure:dos-pure-fe}), where one clearly observes interface C-p states appearing in the band gap from both N and B terminated interfaces. As a result, an almost zero gap situation occurs, which prevents an exponential decay of magnetic interactions.

\subsection{Electronic structure with and without Fe} 

\begin{figure}[ht]
\begin{center}
\includegraphics[scale=0.31]{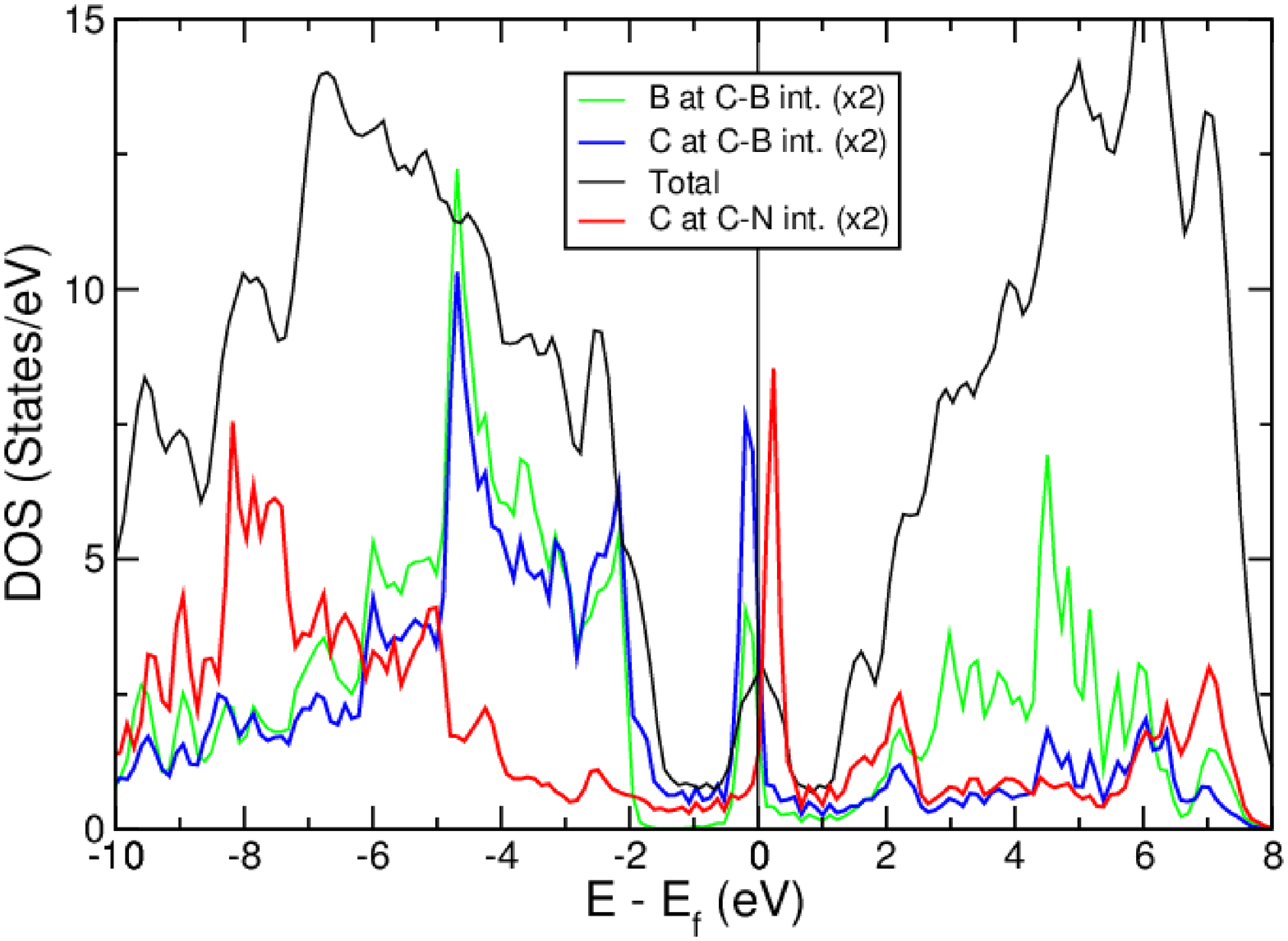}\\
\includegraphics[scale=0.31]{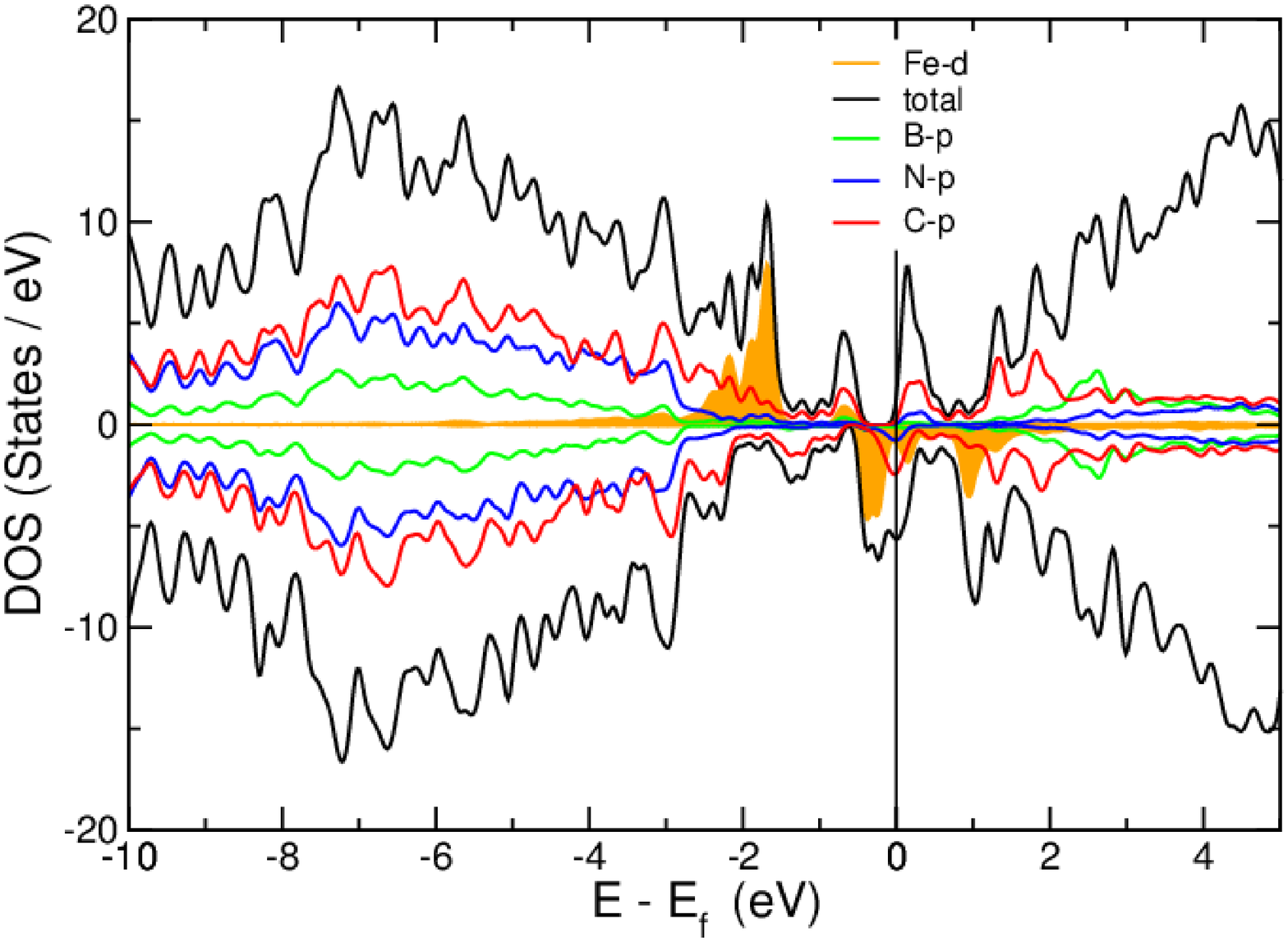}
\end{center}
\caption{(Color online) DOS for (upper panel) pure h-BNC$_2$ sheet and (lower panel) an Fe adatom adsorbed at C-B interface of h-BNC$_2$ sheet. }
\label{figure:dos-pure-fe}
\end{figure}

Since Fe clusters get stuck at the H3 sites, we have studied the densities of states (DOS) of the sheet with one or more Fe$_4$ clusters trapped at H3 sites. In the first
case we calculated the DOS and atom projected DOS (pDOS) for a single Fe adatom 
trapped at a H3 site on a 64-atom h-BNC$_2$ sheet, shown in 
 fig.~\ref{figure:dos-pure-fe}. For comparison, DOS and atom and site projected DOS of 
the pure sheet are also shown. For the sheet without Fe, the valence band states 
just below the gap come mainly from the 
C atoms bonded to B atoms at the C-B interface. Conduction band states just above the gap come mainly from the C atoms at the C-N interface. As mentioned before, states from both C-B and C-N interfaces appear in the gap of pure h-BN. In the presence of Fe, those states still persist, but appear only in the $\downarrow$-spin channel. Along with those, Fe states with d$_{xy}$ and d$_{z^{2}}$ characters occur. The $\uparrow$-spin channel with filled d-states does not contribute at the Fermi energy, yielding a half metallic solution. Fe has a magnetic moment of 2.6 $\mu_{B}$ whereas a few C atoms close to Fe possess small negative moments, yielding a total magnetic moment of 2.0 $\mu_{B}$ in the unit cell. The small DOS at Fermi energy seen in fig.~\ref{figure:dos-pure-fe} occurs due to the use of a broadening parameter.

\begin{figure}[ht]
\begin{center}
\includegraphics[scale=0.33]{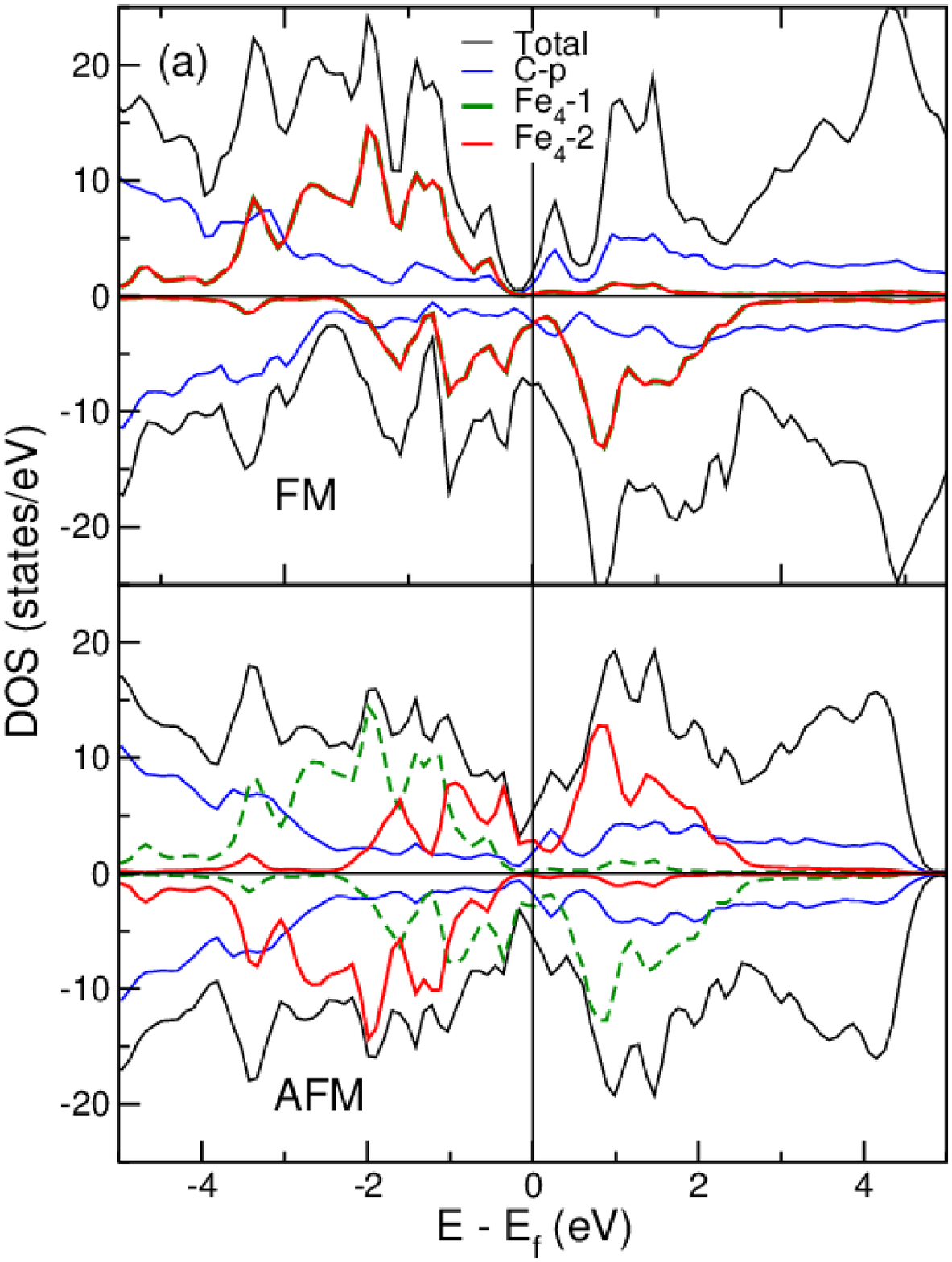}\\
\includegraphics[scale=0.33]{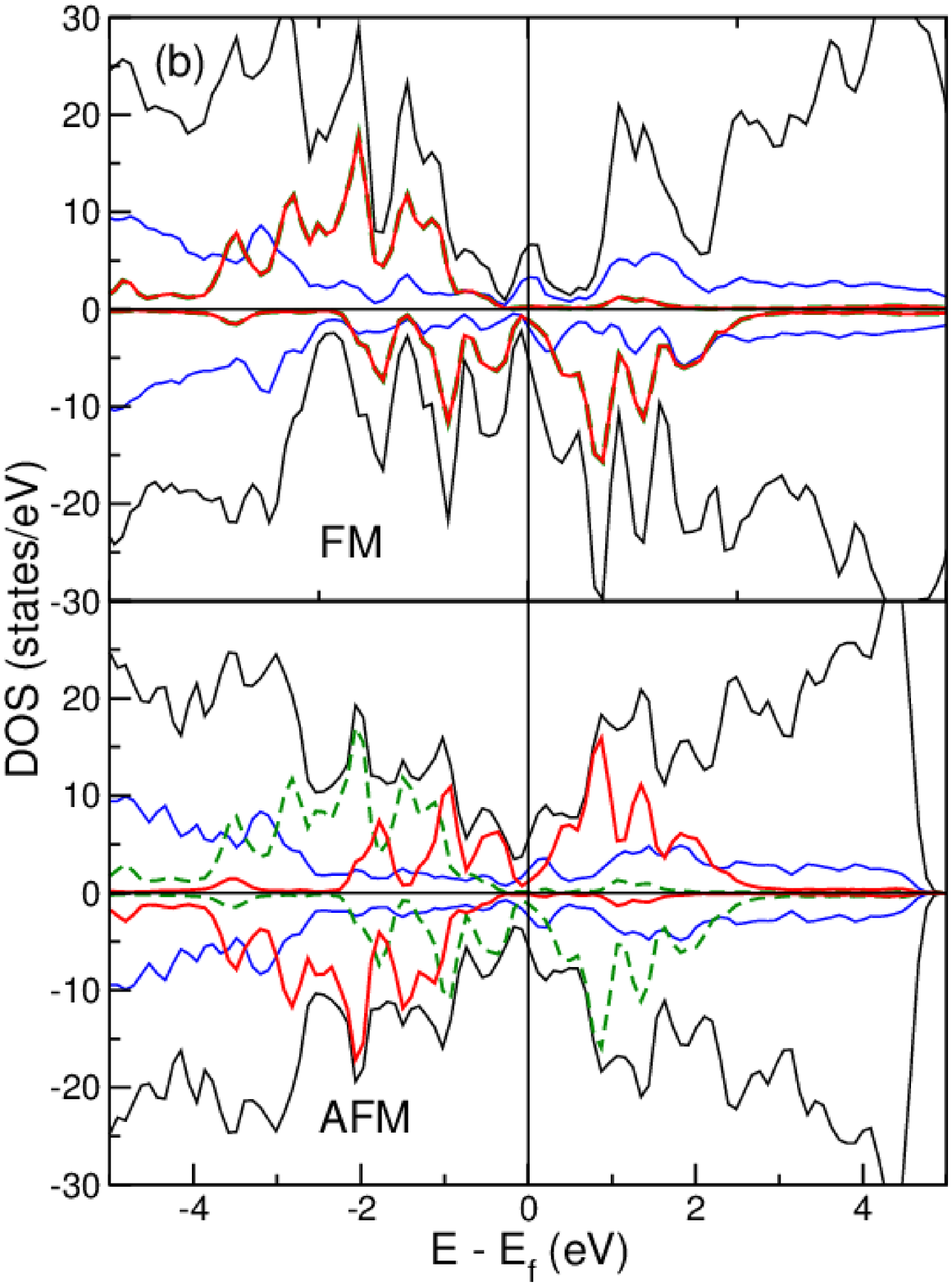}
\end{center}
\caption{(Color online) Density of states of two Fe$_4$ clusters (denoted as Fe$_4$-1 and Fe$_4$-2) adsorbed on the h-BNC$_2$ sheet for (a) 1 row of h-BN and (b) 4 rows of h-BN. For both widths of h-BN, DOSs are shown for FM and AFM alignments between the Fe clusters separated by h-BN.}
\label{figure:dos-exchange}
\end{figure}

In fig.~\ref{figure:dos-exchange}, we show the DOSs of Fe$_4$ clusters separated by h-BN of different widths for both FM and AFM alignments between the Fe clusters' magnetic moments. According to our exchange energy calculations presented in fig.~\ref{figure:exchange}, Fe clusters should couple ferromagnetically (antiferromagnetically) for one row (four rows) wide h-BN domains. The reason for this behavior can be related to the total DOS at the Fermi level (E$_{F}$) for FM and AFM configurations. In  fig.~\ref{figure:dos-exchange} (a), one clearly observes finite total DOS at E$_{F}$ for the AFM configuration in both spin channels whereas, the FM configuration has a vanishing contribution for the $\uparrow$-spin channel. Therefore, the FM configuration is more stable due to lower DOS at E$_{F}$. On the contrary, E$_{F}$ passes through a peak in the DOS for the FM configuration for 4-rows wide h-BN, shown in fig.~\ref{figure:dos-exchange} (b). In this case, the AFM configuration yields lower DOS at E$_{F}$ as it passes through the slope of DOS curves. This makes the AFM configuration more stable than the FM one. 

\subsection{Effect of electron correlation}

So far all the calculations have been performed with the PBE exchange-correlation functional. However, for Fe nanostructures, electron
correlation effects are expected to be important due to narrow band widths. In order to check what effects such electron correlations play in the dynamics and electronic properties of the
Fe clusters on the h-BNC sheet, we have used the PBE+U method \cite{ldau}, where the Coulomb parameter U is added in the Hubbard formalism. For these calculations we used U=4 eV and the intra-atomic exchange parameter J=1 eV. Since there are no experimental
information about the U and J values for Fe clusters deposited on a h-BNC sheet, these results should be treated as a qualitative indication of what
effects electron correlations will have on their properties. As before, we first calculated the
adsorption energies of a single Fe adatom at different hexagonal sites on a 64-atom sheet. The values are given in table ~\ref{table:bind-Fe-ggau}. 
\begin{table}
\caption{Adsorption energies of Fe adatom ($E_a$) at different sites from PBE+U 
calculations. See text for details. The most favorable site H3 is marked in 
bold.}
\label{table:bind-Fe-ggau}
\begin{tabular}{|c|c|c|}
\hline\hline
Position & $E_a$ (eV) & Height (\AA)\\
\hline
H1 & 0.74 & 1.92 \\
H1$'$ & 0.57 & 1.91 \\
H1$"$ & 0.62 & 1.84 \\
H2 & 0.58 & 1.91 \\
{\bf H3} & {\bf 1.11} & {\bf 1.96} \\
H4 & 0.14 & 2.21 \\
\hline
\end{tabular}
\end{table}

It is observed that the adsorption energies decrease at all the sites in presence of a finite U. The heights of the Fe adatom above the sheet concomitantly increase with a decrease in hybridization, but the most stable adsorption site does not change.
The most dramatic effect of electron correlations is seen in the calculated diffusion barriers. With the PBE functional the diffusion barrier between the 
H1$'$ and H1$''$ sites was 1.3 eV. Using PBE+U, the barriers drops to only $\sim$0.12 eV. This is due to the weak hybridization between the adatom and the 2D sheet in presence of strong Coulomb interaction. This raises the question whether the Fe cluster would get trapped
at the C-B interface even when electron correlations are included. To verify this we performed BOMD calculations at $T=300$ K with four Fe adatoms on a
64-atom h-BNC$_2$ sheet using the PBE+U method. Interestingly, the four Fe adatoms do form a cluster as before, and the cluster gets trapped at the C-B interface. Therefore, one can safely conclude that the findings from PBE calculations are qualitatively similar in the case of PBE+U.

\section{Conclusion}
In this paper, we have studied diffusion and magnetism of Fe nanostructures 
(adatom and small clusters) on 2D hybrids of graphene and h-BN by ab initio 
density functional theory. Our calculated diffusion barriers have smaller values 
on the h-BN part compared to those on the graphene part. Ab initio molecular 
dynamics simulations at 300K indicate the formation of Fe clusters from mobile 
Fe adatoms and eventual settlement of the clusters at the C-B interface. 
Moreover, the calculated magnetic exchange couplings between Fe clusters are 
weak and non-monotonous across h-BN regions of varying widths. We conclude that 
the artificially patterned 2D hybrids of graphene and h-BN may act as potential 
substrates for spontaneous formation of magnetic nanostructures at specific 
interfaces.

\section*{Acknowledgement}
SH, BS and OE would like to acknowledge KAW foundation for financial support. In addition, BS acknowledges Carl Tryggers Stiftelse, Swedish Research Council and KOF initiative of Uppsala University for financial support. OE acknowledges support from VR, eSSENCE and the ERC (project 247062 - ASD). We thank SNIC-NSC and SNIC-HPC2N computing centers under Swedish National Infrastructure for Computing (SNIC), and the cluster computing facility at the
Harish-Chandra Research Institute (http://www.cluster.hri.res.in) for granting computer time.

\bibliography{biblioU}

\end{document}